\documentclass[epj]{svjour}
\usepackage{graphics}
\begin{document}
\title{Light-pulse atom interferometry in microgravity}
\author{
    G. Stern\inst{1,3} \and
    B. Battelier\inst{1} \and
    R. Geiger\inst{1} \and
    G. Varoquaux \inst{1} \and
    A. Villing\inst{1} \and
    F. Moron\inst{1} \and
    O. Carraz\inst{2} \and
    N. Zahzam\inst{2} \and
    Y. Bidel\inst{2} \and
    O. Chaibi\inst{3} \and
    F. Pereira Dos Santos\inst{3} \and
    A. Bresson\inst{2} \and
    A. Landragin\inst{3} \and
    P. Bouyer\inst{1}
}
\offprints{}          % Insert a name or remove this line
\institute{Laboratoire Charles Fabry de l'Institut d'Optique, Centre National de la Recherche Scientifique et Universit\'{e} Paris Sud 11, Institut d'Optique Graduate School, RD 128, 91127 Palaiseau Cedex, France
    \and
Office National d'Etude et de Recherches A\'{e}rospatiales, Chemin de la Huni\`{e}re, 91761 Palaiseau, France
    \and
LNE-SYRTE, CNRS UMR 8630, UPMC, Observatoire de Paris, 61 avenue de l'Observatoire, 75014 Paris, France
\date{Received: date / Revised version: date}
\abstract{
We describe the operation of a light pulse interferometer using cold $^{87}$Rb atoms in reduced gravity. Using a series of two Raman transitions induced by light pulses, we have obtained Ramsey fringes in the low gravity environment achieved during parabolic flights. With our compact apparatus, we have operated in a regime which is not  accessible on ground. In the much lower gravity environment and lower vibration level of a satellite, our cold atom interferometer could measure accelerations with a sensitivity orders of magnitude better than the best ground based accelerometers and close to proven spaced-based ones.}
\PACS{
      {PACS-key}{discribing text of that key}   \and
      {PACS-key}{discribing text of that key}
     } % end of PACS codes
} %end of abstract
\maketitle
%
%\section{Introduction}
\label{intro}
Atom interferometry is one of the most promising candidates for ultra-accurate measurements of gravito-inertial forces \cite{dubetsky:2007}, with both fundamental \cite{dimopoulos:2007,dimopoulos:042003,Ertmer:2008,Wolf:2008} and practical (navigation or geodesy) applications. Atom interferometry is most often performed by applying successive \emph{coherent} beam-\-splitting and -\-recombining processes separated by an interrogation time $T$ to a set of particles \cite{borde:1989}. Understanding matter waves interferences phenomena follows from the analogy with optical interferometry \cite{Storey:1994,Borde:2001}: the incoming wave is separated into two wavepackets by a first beam-splitter; each wave then propagates during a time $T$ along a different path and accumulates a different phase; the two wavepackets are finally recombined by a last beam-splitter. To observe the interferences, one measures the two output-channels complementary probability amplitudes which are sine functions of the accumulated phase difference $\Delta \phi$. This phase difference increases with the paths length, i.e. with the time $T$ between the beam-splitting pulses. 

%\section{Experimental setup}
\label{sec:1}

\begin{figure}
\begin{center}
%\begin{minipage}{0.496\columnwidth}
\resizebox{0.96\columnwidth}{!}{%
\includegraphics{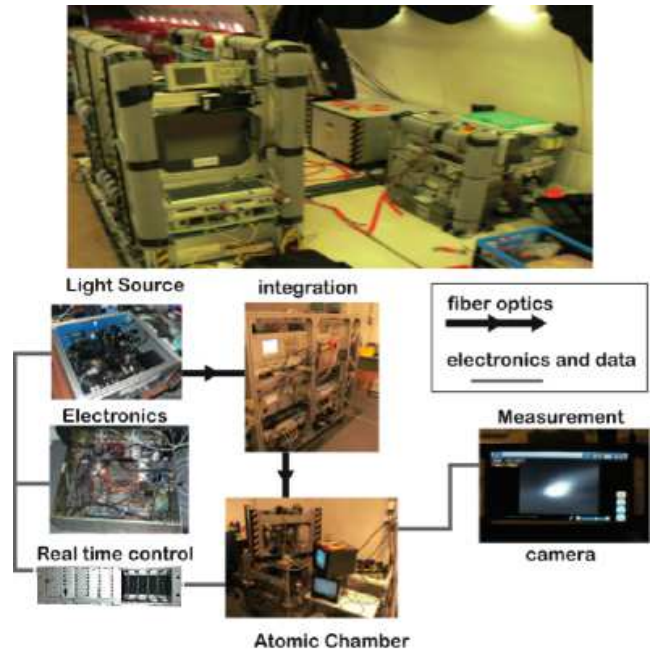}}
\end{center}
\setlength{\unitlength}{\linewidth}

\caption{
Top: The atom interferometer assembled in the Airbus. The main
rack on the left houses the laser sources and the control electronics. The rack
on the front right contains the uninterruptable power-supply, the electrical panel and the high-power laser part. The rack on the back right hosts atom-optics part of the
experiment. Bottom: the architecture of the atom interferometer.
}
\label{fig:photos}
\end{figure}

When used as inertial sensors \cite{Kasevich:1991/2,Gustavson:1997},  the atoms are usually left free to evolve during the interrogation time $T$ so that the interferometer is only sensitive to gravito-inertial effects. In particular, one avoids residual trapping fields that would induce inhomogeneities or fluctuations and would affect the atomic signal. The interrogation time $T$ is consequently limited by, on the one hand, the free expansion of the atomic cloud, and, on the other  hand, the free fall of the atomic cloud. The limitation of expansion is alleviated by the use of ultracold gases \cite{Leanhardt:2003,lecoq:2006}, but, due to free-fall distance, long-interrogation-time experiments require tall vacuum chambers \cite{Clairon:1991}. Laboratory experiments are typically limited to about 300 ms of free fall with a 1 m-tall apparatus if the atoms are simply released, or more by launching them upward as in atomic fountains. This can be increased on much larger apparatuses : a 10 m-high atom interferometer is currently under construction at Stanford \cite{dimopoulos:042003}, giving access to 1.4 seconds of interrogation time. Free-fall heights of more than 100 m, corresponding to durations of about 5 seconds are also available in a drop tower (ZARM Bremen, Germany  \cite{vogel:2006}). Another solution consists in  performing the experiments in microgravity achieved during parabolic flights provided by an aeroplane, as for the PHARAO prototype \cite{laurent:1998}.  In the I.C.E. (\textit{Interf\'erometrie Coh\'erente pour l'Espace}) collaboration \cite{nyman:2006,varoquaux:2007} that we present here, we are conducting cold-atom interferometry experiments in such an airplane (the A-300 0-G Airbus of NOVESPACE), which carries out ballistic flights. Microgravity is obtained via 20 seconds-long parabolas by steering the plane to cancel drag and follow free fall. The residual acceleration is on the order of $10^{-2}$g. With 90 parabolas per flight session, we have access to 30 minutes total of reduced gravity. In this letter, we present a first validation of our 0-g setup by obtaining Ramsey fringes with copropagating Raman transitions during parabolas, the interrogation times being longer than those we could obtain on Earth with the same configuration. 

Transferring a laboratory-bound cold atoms interferometer into an automated experiment suitable for microgravity use poses many technical challenges \cite{vogel:2006,nyman:2006,KAnemann:2007hl,yu:2004}. We assembled a prototype atomic source suitable for inertial-sensing in an airplane from the I.C.E. collaboration components \cite{nyman:2006} (see Fig. \ref{fig:photos}). The atom interferometer is made of 4 elements: a vacuum chamber with optics; lasers sources for cooling and coherent manipulation of atoms; a stable oscillator (in our case a hyperfrequency source at about 6.8 GHz \cite{nyman:2006}) which is a frequency reference for the Raman lasers; and an autonomous real-time controller for the experimental sequence and data calculations.  For the interferometric measurement, we prepare clouds of cold $^{87}$Rb in a Magneto-Optical Trap (MOT) and release them for interrogation during their free fall.

\begin{figure}[b]
\begin{center}
\resizebox{0.96\columnwidth}{!}{
%\rotatebox[origin=rB]{270}{
\includegraphics{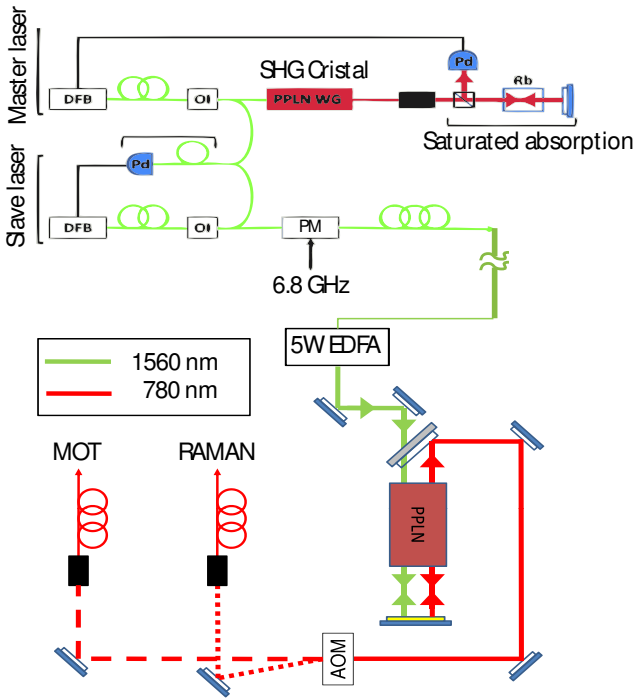}}%}
\end{center}
\caption{
Diagram of the laser system. The master laser (linewidth of $1$ MHz) is
a monolithic semiconductor element: a $1560$ nm Distributed 
Feed-Back (DFB) fiber laser is frequency doubled in a PPLN waveguide; the resulting
$780$ nm light is then sent into a saturated-absorption
spectroscopy setup for frequency locking on a $^{85}$Rb transition; the slave is a $80$ mW DFB laser diode at $1560$ nm and is frequency-locked
on the master laser by measuring the frequency of their beat-note recorded on a fibered fast photodiode. Frequency control of the lasers is achieved via feedback to their supply current. After amplification through a 5W-Erbium-Doped Fiber
Amplifier (EDFA), the slave laser is frequency doubled in free space with a bulk PPLN crystal; we obtain about
0.3 W at $780$ nm.}
\label{fig:laser}
\end{figure}

Moving away from extended-cavity-laser-diode-based systems, as developed in the PHARAO project \cite{laurent:2006}, we have designed
laser sources at $780$ nm for cooling and coherently manipulating the atoms that rely on telecom technologies and second harmonic generation \cite{Bruner:1998,Thompson:2003}. This allows to use fiber-optics components and offers a reliable, robust and compact system, quite insensitive to the environmental perturbations encountered in the airplane. These novel laser sources are very similar to the ones described in
details in \cite{lienhart:2007} so we limit here to outlining the successful design. A first reference DFB 1560 nm pigtailed laser diode (linewidth $\sim$ 1 MHz) is frequency doubled in a PPLN waveguide and locked on a $^{85}$Rb transition through a saturated absorption setup (see Fig. \ref{fig:laser}). A slave DFB 1560 nm pigtailed diode, similar to the first one, is  locked to the first laser at a frequency difference monitored through the beat note signal, as measured by a fibered fast photodiode. The frequency offset can be adjusted so that the slave DFB is red detuned from the resonance of the $F=2\rightarrow F=3$ transition of $^{87}$Rb with a detuning ranging from 0 to 1.1 GHz. A 1560 nm fibered phase modulator is then used to generate two sidebands $\sim$ 6.8 GHz apart. One of these sidebands acts either as the repumping laser during the cooling phase, or as the second Raman laser during coherent manipulation of the atoms, depending on the applied frequency.  

\begin{figure}
\begin{center}
%\begin{minipage}{0.496\columnwidth}
\resizebox{0.96\columnwidth}{!}{%
\includegraphics{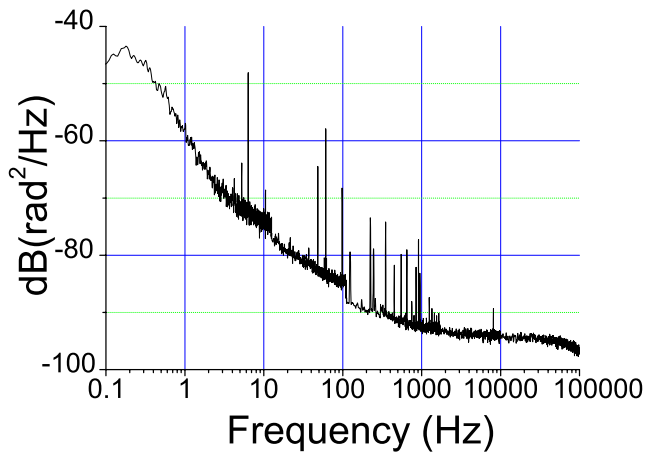}}
\end{center}
\setlength{\unitlength}{\linewidth}

\caption{Spectral density of the phase noise of the quartz recorded by comparison with other ultra-stable quartz oscillators phase-locked on H-Maser of the SYRTE.}
\label{fig:spectre}
\end{figure}

The microwave reference has been simplified compared to \cite{nyman:2006} in order to make it more reliable in the plane environment. It's based on a direct multiplication of a 10 MHz quartz oscillator to 6.8 GHz without any intermediate oscillator or phase lock loop. The ultra-stable quartz has been chosen to be a good compromise to achieve low phase noise at low and high frequencies simultaneously (see Fig. \ref{fig:spectre}), as in \cite{laurent:1998}. The multiplication is done in three steps: a first multiplication by 10 to 100 MHz (commercial Wenzel system), then multiplication by 2 and finally to 6.8 GHz by a comb generator (non-linear transmission line, Wenzel model 7100). Two direct digital synthesis (DDS) are used to adjust the cooling/repumping frequency difference and the Raman beams frequency difference respectively.

%\section{Experimental setup}
\label{sec:2}

After amplification through a 5W Erbium-Doped Fiber Amplifier (EDFA), the slave laser is frequency doubled in free space with a double-pass in a 4 cm bulk PPLN crystal. We typically obtain $\sim$ 300 mW at 780 nm. A 80 MHz Acousto-Optical Modulator (AOM) is used to switch between the MOT configuration (in which the non diffracted order of the AOM is used) and the Raman configuration (in which the first diffraction order is used, see Fig. \ref{fig:laser}). The use of the first order of the AOM for the Raman beam enables to create ultra-short pulses of light (10 $\mu$s typically). Additional mechanical shutters ensure a total extinction of the beams. Two optical fibers finally bring the MOT and the Raman beams to the science chamber. 

The fibers deliver the light to the vacuum-chamber module \cite{varoquaux:2007}. The MOT fiber is sent to a 1-to-3 fiber beam-splitter \footnote{From Sch\"after und Kirchhoff : \texttt{http://www.sukhamburg.de/}}  
which delivers three beams which are then retroreflected and produce the MOT. The circularly-polarized Raman beam has a 1 inch diameter  and is aligned with the horizontal plane. A 300 mG horizontal magnetic field is aligned with the Raman beam to raise the Zeeman degeneracy  of the hyperfine sub-levels. The intensity of the lasers can be up to 20 times the saturation intensity of rubidium, which
allows for short Raman pulses with weak velocity selection. The Raman detuning is about 700 MHz. The effective Rabi pulsation $\Omega_{\rm eff}$ is about $2\pi\times12.5$ kHz. Finally, a magnetic shield around the science chamber prevents from changes of the Earth's magnetic field directions during parabolas \cite{varoquaux:2007}. 

%\section{Optical Ramsey fringes in microgravity}

\begin{figure}
\begin{center}
%\begin{minipage}{0.496\columnwidth}
\resizebox{0.96\columnwidth}{!}{%
\includegraphics{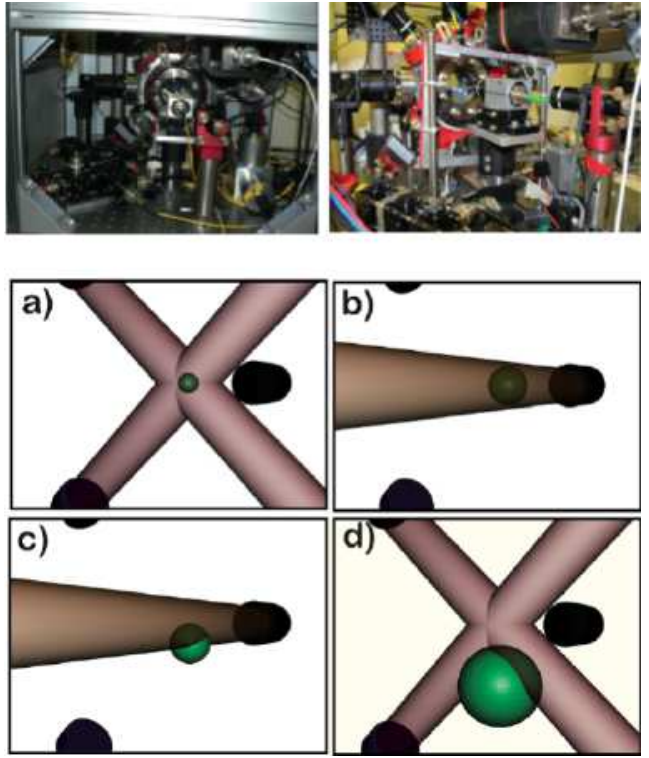}}
%\end{minipage}
\end{center}
\setlength{\unitlength}{\linewidth}
\caption{
Up: inside the atom-optics rack (the vacuum chamber and the free-space
optics). Down : 3D schematics showing the beams configuration and the interferometer sequence. 
a) Atom cooling and trapping (MOT, the horizontal retroreflected MOT beam is not shown for clarity), b) First Raman interrogation pulse colinear to the horizontal retroreflected MOT beam, c) second Raman interrogation pulse and d) detection with the MOT beams. The atomic cloud is represented as falling under gravity.}
\label{fig:schematics}
\end{figure}

\begin{figure*}
\begin{center}
\resizebox{1.9\columnwidth}{!}{%
\includegraphics{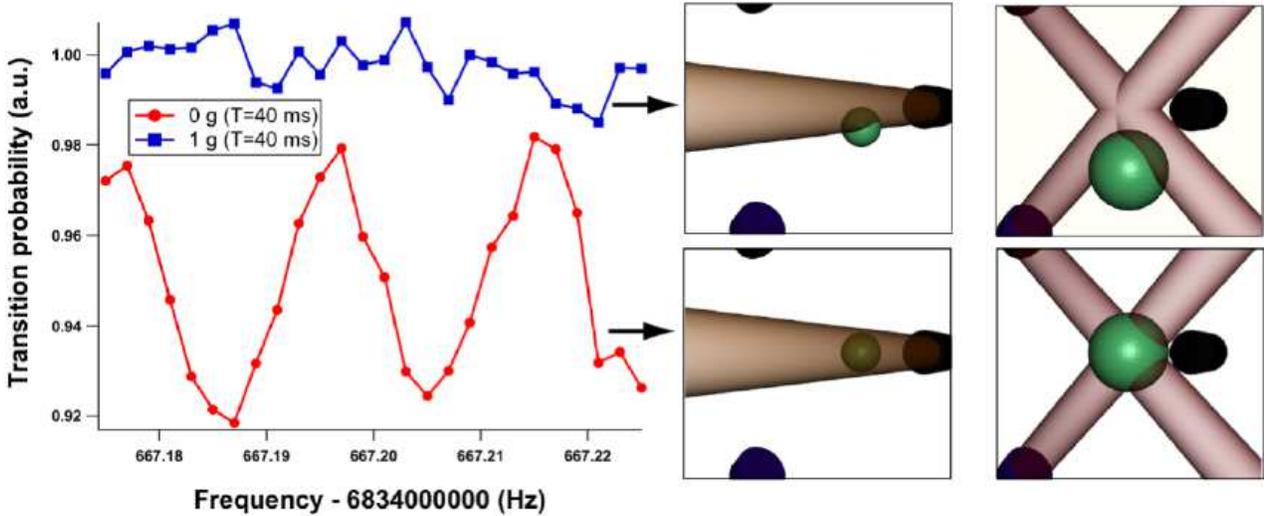}}
\end{center}
\caption{Comparison of fringes with or without gravity. 
In the presence of gravity (upper curve and pictures), the atoms, when released from the trap
fall down, and eventually exit the Raman pulse region and the recapture volume. The atomic signal then drops and no fringes are visible. In the absence of gravity (lower curve and pictures), even for longer times, the atoms stay in the 
Raman pulse region and recapture area. The detection limit is eventually set by the temperature induced expansion of the atomic cloud. }
\label{fig:fringe}
\end{figure*}

The science chamber in which we operate our atom interferometer is shown in Fig. \ref{fig:schematics}. We load about 10$^9$ atoms in the MOT from a Rubidium vapor in 500 ms. We release the atoms from the MOT and further cool them down below 100 $\mu$K during a brief phase of optical molasses. Then, we prepare the atoms in the lower hyperfine state $F=1$ using optical pumping. After the extinction of the MOT beams, we shine the atoms with two Raman light pulses separated by a time $T$. The duration $\tau$ of these pulses is chosen such that $\Omega_{\rm eff}\times\tau=\frac{\pi}{2}$ (splitting of the matter wave). The Raman lasers are copropagating so that a nearly zero momentum is transfered during the Raman transition. In this configuration, the two successive $\frac{\pi}{2}$ pulses enable to record optically induced Ramsey fringes that are the signature of the matter waves interferences between the two interferometer paths \cite{Antoine:2003/2,Borde:2002}. After the Raman pulses, the MOT beams are switched on at resonance and a photodiode monitors the fluorescence which is proportional to the number of atoms. During a few milliseconds, the microwave source is first turned off (absence of the repumping laser) to record the number $N_2$ of atoms in $F=2$. Second, we switch on the microwave source (presence of the repumping sideband) and the photodiode detects the total number of atoms  $N$. Plotting the ratio $N_2/N$ with respect to the frequency of the Raman transitions sideband, we thus obtain Ramsey fringes, corresponding to the proportion of atoms having undergone coherent transfer between the two states $|F=1, m_F=0\rangle$ and $|F=2, m_F=0\rangle$ (see Fig. \ref{fig:fringe}). 

\begin{figure}
\begin{center}
\resizebox{.8\columnwidth}{!}{%
\includegraphics{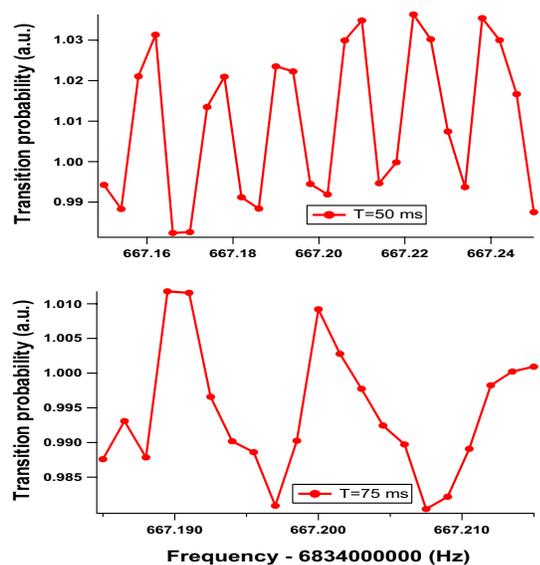}}
\end{center}
\caption{
Optical Ramsey fringes obtained in micro gravity for $T=50$ ms and $T=75$ ms, (corresponding to respectively 18 mm and 35 mm of free fall under gravity). The x-axis is the frequency of the Raman transition sideband delivered by the hyperfrequency source and the y-axis represents the non-normalized ratio $N_2/N$.}
\label{fig:fringelong}
\end{figure}

The total duration of the sequence is about $T+10$ ms from the end of the molasses phase until the detection. In the lab, i.e. in a 1-g environment, we can typically detect atoms until $T=20$ ms. Above this limit, the free fall of the atoms is too important and the atoms exit both the area of the Raman beams and the detection volume, as shown in the schematics of figure \ref{fig:fringe}. However, $T$ can be much longer during parabolas where the residual acceleration is on the order of $10^{-2}$ g.  Figure \ref{fig:fringe} clearly illustrates the advantage of such a reduced gravity environment: with $T=40$ ms, we have recorded fringes which could not be observed on ground; we could operate at a largest pulse time interval of $T=75$ ms (see Fig. \ref{fig:fringelong}) which represents a fringe period of $1/T\sim$ 14 Hz. We are prone to believe that the main limitation for longer values of $T$ is due to the cloud temperature: the spatial extension of the cloud increases, and the Rabi frequency is then not the same for all the atoms, depending on the laser intensity at their position. It can explain the decrease of the fringes amplitude when T is longer. This effect is enhanced by the residual acceleration -  $4\times10^{-2}$g during about 100 ms leads to motion amplitudes of 2 mm which is enough to reduce the Raman beams efficiency. Atoms also exit the detection area, and this makes the signal-to-noise ratio drop.

%\section{Conclusion}
The vibration noise in the plane do not enable us to use the Raman beam in a velocity selective configuration yet \cite{Kasevich:1991}, and thus we could not render this interferometer sensitive to inertial effects. Consequently, the residual acceleration noise has been measured with accelerometers locally anchored to the experimental apparatus. This noise corresponds to large residual accelerations ($\sim$ 10 cm/s$^2$) that will doppler shift the resonance and thus hinder it. Different techniques can be used to reduce the influence of these spurious accelerations: active stabilization of the retroreflecting mirror \cite{Hensley:1999}, post-corrections or feed-forward from an accelerometer signal on the Raman phase \cite{LeGouet:2008},  or combination of vibrations measurements by a seismometer and the measured transition probabilities \cite{Merlet:2009}. With the use of an appropriate vibration isolation, it will allow interrogation of the freely-falling atoms during several  seconds, and reach high-precision not yet achieved with ground based atom inertial sensors. When implemented in space, with a residual noise lower than $10^{-6}$ g, the sensibility can reach that of the best spaced-based accelerometers\cite{Odyssey}. 

To conclude, we have successfully tested a cold atom light pulse interferometer in
aircraft parabolic flights. Our preliminary results show that laboratory experiments can be adapted for this new 
experimental platform and used to develop the future generation of air/spaceborn atom inertial sensors. Our experimental set-up offers an unprecedented platform for development of future fundamental physics instruments to test General Relativity of gravitation. 
Unlike orbital platforms, development cycles on ground-based facilities (either in a plane or in a drop tower) can be  short enough to offer rapid technological evolution for these future sensors.  In the future, high-precision drag-free space-born applications will require further progress to achieve longer interrogation times using ultra-low velocity atoms. New-generation of degenerate atomic source design are currently under study for that purpose \cite{vogel:2006,nyman:2006}.
 
\section*{Acknowledgments}

The I.C.E. collaboration is funded by the Centre National d'Etudes Spatiales, as are GS, and RG. This work is also supported by the European Space Agency under the "Space Atom Interferometry" program. Further support comes from the RTRA "triangle de la physique" and the European Union STREP consortium FINAQS. Laboratoire Charles Fabry, ONERA and SYRTE are all members of IFRAF\footnote{Institut Francilien de Recherche en Atomes Froids : \texttt{http://213.251.135.217/ifraf/}}.

\bibliographystyle{epj}
\bibliography{biblioFin}

\begin{thebibliography}{30}

\bibitem{dubetsky:2007}
B.~Dubetsky, M.A. Kasevich, Physical Review A (Atomic, Molecular, and Optical
  Physics) \textbf{74}(2), 023615 (~17) (2006),
  \texttt{http://link.aps.org/abstract/PRA/v74/e023615}

\bibitem{dimopoulos:2007}
S.~Dimopoulos, P.W. Graham, J.M. Hogan, M.A. Kasevich, Phys. Rev. Lett.
  \textbf{98}(11), 111102 (~4) (2007)

\bibitem{dimopoulos:042003}
S.~Dimopoulos, P.W. Graham, J.M. Hogan, M.A. Kasevich, Physical Review D
  (Particles, Fields, Gravitation, and Cosmology) \textbf{78}(4), 042003 (~29)
  (2008), \texttt{http://link.aps.org/abstract/PRD/v78/e042003}

\bibitem{Ertmer:2008}
W.~Ertmer, C.~Schubert, T.~Wendrich, M.~Gilowski, M.~Zaiser, T.~v.~Zoest,
  E.~Rasel, C.J. Bord{\'e}, A.~Clairon, P.L. et~al., Experimental Astronomy
  \textbf{23}(2), 611 (2009)

\bibitem{Wolf:2008}
P.~Wolf, C.J. Bord{\'e}, A.~Clairon, L.~Duchayne, A.~Landragin, P.~Lemonde,
  G.~Santarelli, W.~Ertmer, E.~Rasel, F.S.C. et~al., Exp Astron \textbf{23},
  651 (2009)

\bibitem{borde:1989}
C.J. Bord{\'e}, Phys. Lett. A  (1989)

\bibitem{Storey:1994}
P.~Storey, C.~Cohen-Tannoudji, J. Phys. II France \textbf{4}, 1999 (1994)

\bibitem{Borde:2001}
C.J. Bord\'e, C. R Acad. Sc. - s{\'e}ries IV - Physics \textbf{2}(3), 509
  (2001)

\bibitem{Kasevich:1991/2}
M.~Kasevich, S.~Chu, Phys. Rev. Lett. \textbf{67}, 181 (1991)

\bibitem{Gustavson:1997}
T.L. Gustavson, P.~Bouyer, M.A. Kasevich, Phys. Rev. Lett. \textbf{78}, 2046
  (1997)

\bibitem{Leanhardt:2003}
A.~Leanhardt, T.~Pasquini, M.~Saba, A.~Schirotzek, Y.~Shin, D.~Kielpinski,
  D.~Pritchard, W.~Ketterle, Science \textbf{301}, 1513 (2003)

\bibitem{lecoq:2006}
Y.~Le~Coq, J.~Retter, S.~Richard, A.~Aspect, P.~Bouyer, App. Phys. B
  \textbf{84}(4), 627 (2006)

\bibitem{Clairon:1991}
A.~Clairon, C.~Salomon, S.~Guelatti, W.~Phillips, Europhys. Lett. \textbf{16},
  165 (1991)

\bibitem{vogel:2006}
A.~{Vogel}, M.~{Schmidt}, K.~{Sengstock}, K.~{Bongs}, W.~{Lewoczko},
  T.~{Schuldt}, A.~{Peters}, T.~{van Zoest}, W.~{Ertmer}, E.~{Rasel} et~al.,
  Appl. Phys. B \textbf{84}, 663 (2006)

\bibitem{laurent:1998}
P.~{Laurent}, P.~{Lemonde}, E.~{Simon}, G.~{Santarelli}, A.~{Clairon},
  N.~{Dimarcq}, P.~{Petit}, C.~{Audoin}, C.~{Salomon}, European Physical
  Journal D \textbf{3}, 201 (1998)

\bibitem{nyman:2006}
R.A. {Nyman}, G.~{Varoquaux}, F.~{Lienhart}, D.~{Chambon}, S.~{Boussen}, J.F.
  {Cl{\'e}ment}, T.~{M{\"u}ller}, G.~{Santarelli}, F.~{Pereira Dos Santos},
  A.~{Clairon} et~al., Appl. Phys. B \textbf{84}, 673 (2006)

\bibitem{varoquaux:2007}
G.~{Varoquaux}, N.~{Zahzam}, W.~{Chaibi}, J.F. {Cl{\'e}ment}, O.~{Carraz}, J.P.
  {Brantut}, R.A. {Nyman}, F.~{Pereira Dos Santos}, L.~{Mondin}, M.~{Rouz{\'e}}
  et~al., \emph{{I.C.E.: An Ultra-Cold Atom Source for Long-Baseline
  Interferometric Inertial Sensors in Reduced Gravity}}, in \emph{Proceedings
  of the XLIInd Rencontres de Moriond, Gravitational Waves and Experimental
  Gravity}, edited by J.~Dumarchez, J.T.T. V\^an (2007), p. 335,
  \texttt{arXiv:physics/0705.2922}

\bibitem{KAnemann:2007hl}
T.~K\"onemann, W.~Brinkmann, E.~G\"ok\"u, C.~L\"ammerzahl, H.~Dittus, T.~van
  Zoest, E.M. Rasel, W.~Ertmer, W.~Lewoczko-Adamczyk, M.~Schiemangk et~al.,
  Applied Physics B: Lasers and Optics \textbf{89}(4), 431 (2007),
  \texttt{http://dx.doi.org/10.1007/s00340-007-2863-8}

\bibitem{yu:2004}
N.~{Yu}, J.M. Kohel, J.R. Kellogg, L.~Maleki, App. Phys. B \textbf{84}(4), 647
  (2006)

\bibitem{laurent:2006}
P.~{Laurent}, M.~{Abgrall}, C.~{Jentsch}, P.~{Lemonde}, G.~{Santarelli},
  A.~{Clairon}, I.~{Maksimovic}, S.~{Bize}, C.~{Salomon}, D.~{Blonde} et~al.,
  App. Phys. B \textbf{84}, 683 (2006)

\bibitem{Bruner:1998}
A.~Bruner, V.~Mahal, I.~Kiryuschev, A.~Arie, M.A. Arbore, M.M. Fejer, Applied
  Optics \textbf{37}(27), 6410 (1998)

\bibitem{Thompson:2003}
R.J. Thompson, M.~Tu, D.C. Aveline, N.~Lundblad, L.~Maleki, Optics Express
  \textbf{11}(14), 1709 (2003)

\bibitem{lienhart:2007}
F.~Lienhart, S.~Boussen, O.~Carraz, N.~Zahzam, Y.~Bidel, A.~Bresson, Applied
  Physics B: Lasers and Optics \textbf{89}(2), 177 (2007),
  \texttt{http://dx.doi.org/10.1007/s00340-007-2775-7}

\bibitem{Antoine:2003/2}
C.~Antoine, C.J. Bord{\'e}, J. Opt. B : Quantum. Semiclass. Opt \textbf{5},
  S199 (2003)

\bibitem{Borde:2002}
C.J. Bord{\'e}, Metrologia \textbf{39}, 435 (2002)

\bibitem{Kasevich:1991}
M.~Kasevich, D.S. Weiss, E.~Riis, K.~Moler, S.~Kasapi, S.~Chu, Phys. Rev. Lett.
  \textbf{66}(18), 2297 (1991)

\bibitem{Hensley:1999}
J.~Hensley, A.~Peters, S.~Chu, Review of Scientific Instruments \textbf{70
  (6)}, 2735 (1999)

\bibitem{LeGouet:2008}
J.L. Gou{\"e}t, T.~Mehlst{\"a}ubler, J.~Kim, S.~Merlet, A.~Clairon,
  A.~Landragin, F.~{Pereira Dos Santos}, Applied Physics B: Lasers and Optics
  \textbf{92} (2008)

\bibitem{Merlet:2009}
S.~Merlet, J.L. Gou{\"e}t, Q.~Bodart, A.~Clairon, A.~Landragin, F.~{Pereira Dos
  Santos}, P.~Rouchon, Metrologia \textbf{46}, 87 (2009)

\bibitem{Odyssey}
B.~Christophe, P.~Andersen, J.~Anderson, S.~Asmar, P.~B\'erio, O.~Bertolami,
  R.~Bingham, F.~Bondu, P.~Bouyer, S.~Bremer et~al., Experimental Astronomy
  \textbf{23}(2), 529 (2009),
  \texttt{http://dx.doi.org/10.1007/s10686-008-9084-y}

\end{thebibliography}

\end{document}